\documentclass{ws-p8-50x6-00}

\begin{document}

\title{Analogies, new paradigms and observational data as growing factors of Relativistic Astrophysics}

\author{Remo J. Ruffini} 

\address{Physics Department 
  and \\
  International Center for Relativistic Astrophysics, 
  University of Rome, I--00185 Roma, Italy}


\maketitle{Patterns in the scientific developments of Relativistic Astrophysics are analyzed with special attention to the physics and astrophysics of Black Holes and Gamma Ray Bursts.}

\section{Introduction}
 Hagen Kleinert has been a pioneer in establishing analogies among widely separated fields of theoretical physics applying relativistic quantum field theories techniques, notoriously  his classical work with Richard Feynman \cite{feynman} , to the treatment of a variety of topics of research ranging  from condensed matter, to crystal melting, polymer physics, phase transitions, differential geometry  etc.  I have always been very impressed by his profound knowledge of physics and his courage in approaching a so vast program of research as testified also in his classical books \cite{kleinert1,kleinert2,kleinert3}. In the occasion of his sixtieth birthday I am dedicating to him a few brief considerations, based on my work, of how the role of analogies among different fields, the establishment of new paradigms as well as the crucial and timely arrival of observational data have signed the development of Relativistic Astrophysics.
\section{The birth of Relativistic Astrophysics}
In 1931, in analogy with the developments made  in the field of atomic physics by Enrico Fermi and his school \cite{fermi}, Lev  D. Landau \cite{landau} was lead to introduce a new paradigm in the approach of astrophysical problems, not based on hypothesis chosen just by mathematical convenience but grounded on the new concepts developed in theoretical physics.. The use of the Fermi-Dirac statistics to study the equilibrium of a star paved the way to analyze the latest phases of the evolution of a star.. This  program was developed in an entire new paradigm for the understanding of white dwarfs in the classical book by Chandrasekhar\cite{chandra0}.Similarly new paradigms were introduced, using the general relativistic techniques by E. Tolman  and the seminal ideas of George Gamow, by Robert Julius Oppenheimer: I recall here the treatment of neutron stars in the classic works with his student G.M. Volkoff \cite{oppen}. 
This entire field of research reached full maturity with the discovery of pulsars in 1968 and especially with  the discovery of the pulsar in the crab nebula. The observation of the period of that pulsar and his slowing down rate  clearly pointed to unequivocal evidence for the identification of the first neutron star in the galaxy but also to the understanding that the energy source  of pulsars were very simply the rotational energy of the neutron star. The year 1968 can be definitely considered the birthdate of Relativistic Astrophysics.

I was in Princeton in those days initially as a  postdoctoral fellow at the University in the group of John Archibald Wheeler, then as a Member of the Institute for Advanced Study and then as an instructor and assistant professor at the University. The excitement for the neutron stars discovery  boldly led us directly to a yet unexplored classic paper by Robert Julius Oppenheimer and Snyder "on continued gravitational contraction" \cite{snyder} and this opened up an entire new field of research to which I have dedicated all the rest of my life and is giving, still today, some distinctively important results. I will comment in the following a few crucial moments, the way I remember them, that influenced very much the development of Relativistic Astrophysics with particular emphasis on the establishment of analogies, new paradigm and crucial observational data.

\section{Analogies between trajectories of Cosmic Rays and trajectories in General relativity}
\begin{figure}
\vspace{-.5cm}
\epsfxsize=6.0cm
\begin{center}
\mbox{\epsfbox{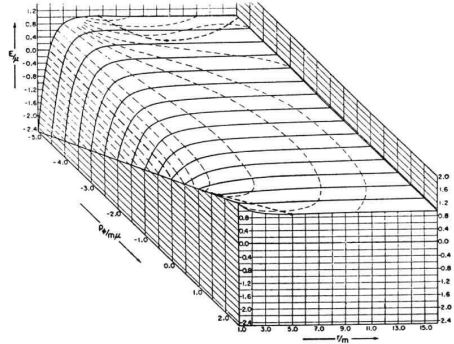}}
\end{center}
\vspace{-0.2cm}
\caption[]{``Effective potential" around a Kerr black hole, see Ruffini and Wheeler 1971}
\label{poten}
\end{figure}

Analogies between trajectories of Cosmic Rays and trajectories in General relativity
An ``effective
potential" technique had been used very successfully by Carl St\o rmer in the
1930s in studying the trajectories of cosmic rays in the Earth's magnetic field
(St\o rmer 1934) \cite{s34}.
In the fall of 1967 Brandon Carter visited Princeton and presented his
remarkable mathematical work leading to the separability of the
Hamilton-Jacobi equations for the trajectories of charged particles in the
field of a Kerr-Newmann geometry (Carter 1968) \cite{bc}. This visit had a profound impact on our small
group working with John Wheeler on the physics of gravitational collapse.
Indeed it was Johnny who had the idea to exploit the analogy between  the trajectories of Cosmic Rays and the trajectories in General relativity, using the St\o rmer ``effective
potential" technique in order to obtain physical consequences from the
set of first order differential equations obtained by Carter. I still remember the $2m\times 2m$ grid plot of the effective potential for particles
around a Kerr metric I prepared which finally appeared in print (Rees, Ruffini and Wheeler 1973,1974 \cite{rrw}; see Fig.(\ref{poten}). 
From this work came the celebrated result of the maximum binding energy of  $1 - {1 \over \sqrt{3}}\sim42\%$ 
for corotating orbits and 
$1-{5\over 3\sqrt{3}}\sim 3.78\%$ for counter-rotating orbits
 in the Kerr geometry. We were very pleased to be associated with Brandon Carter in a ``gold
medal" award for this work presented by Yevgeny Lifshitz: in the last edition of volume 2 of the Landau
and Lifshitz series ({\it The Classical Theory of Fields\/}), both Brandon's work and my own work with Wheeler were proposed as named exercises for bright students!
This was certainly a simple and fruitful analogy which led to a successful theoretical accomplishment, opening up a new window on very unexpected general relativistic effects. It is interesting that in recent years presented it has become clear that the difference in the binding energies of the corotating and counterrotating orbits in a Kerr Newmann geometry have become the object of direct astrophysical observations in binary X-Ray sources.

\section{Analogy  of the "Black Hole" with an elementary physical system}
\begin{figure}
\vspace{-.5cm}
\epsfxsize=6.0cm
\begin{center}
\mbox{\epsfbox{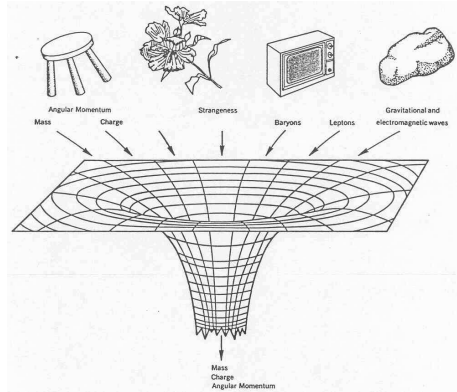}}
\end{center}
\vspace{-0.2cm}
\caption[]{The black hole uniqueness theorem.}
\label{tvsetbh}
\end{figure}

In our article ``Introducing the Black Hole" (Ruffini and Wheeler 1971) \cite{rw71} we first proposed the famous "uniqueness theorem" , stating that black holes can only be characterized
by their mass-energy $E$, charge $Q$ and angular momentum $L$. This analogy, between a Black Hole and a most elementary physical system,  was immaginifically represented by Johnny in a  very unconventional figure in which TV sets, bread, flowers and other objects lose their characteristic features and merge in the process of gravitational collapse into the three fundamental parameters of a black hole, see Fig.~\ref{tvsetbh}. That picture became the object of a great deal of lighthearted discussion in the physics community. A proof of this uniqueness theorem, satisfactory for some case of astrophysical interest, has been obtained after twenty five years of meticulous mathematical work (see e.g., Regge and Wheeler \cite{ReggeW}, Zerilli \cite{Zerilli1,Zerilli2}, Teukolsky \cite{teukolsky} , C.H. Lee   \cite{lee}, Chandrasekhar \cite{chandra}). However the proof  still presents some outstanding difficulties in
its most general form. Possibly some progress will be reached in the near future
with the help of computer algebraic manipulation techniques to overcome the
extremely difficult mathematical calculations (see e.g., Cruciani (1999) \cite{cru}, Cherubini and Ruffini  (2000) \cite{chrr}
 Bini et al.  (2001) \cite{bcjr1}, Bini et al.  (2001) \cite{bcjr2}).

It is interesting that this proposed analogy, which appeared at first to be almost trivial, has revealed to be one of the most difficult to be proved implying a monumental work, unsurpassed in difficulty, both in mathematical physics and relativistic field theories. The fact that this analogy is still unproven and that the most general perturbation of a Black Hole endowed with electromagnetic structure (EMBH) and rotation is still far from being solved offers an extremely good example of the difference between General Relativity and classical physics. The solution of this problem from a mathematicalphysics point of view may have profound implications on our understanding of fundamental physical laws.

\section{Analogy between pulsars and Black Hole Physics; the extraction of rotational energy}

We were still under the sobering effects of the pulsar discovery and the very clear explanation by Tommy Gold and Arrigo  Finzi that the rotational energy of the neutron star had to be the energy source of the pulsars phenomenon, when in 1969 the first meeting of the European
Physical Society in Florence in 1969. In a splendid talk Roger Penrose \cite{p69} advanced the possibility that, much like in Pulsars, also , analogously, in the case of black holes the rotational energy could be in principle  extracted.
\begin{figure}
\vspace{-.5cm}
\epsfxsize=6.0cm
\begin{center}
\mbox{\epsfbox{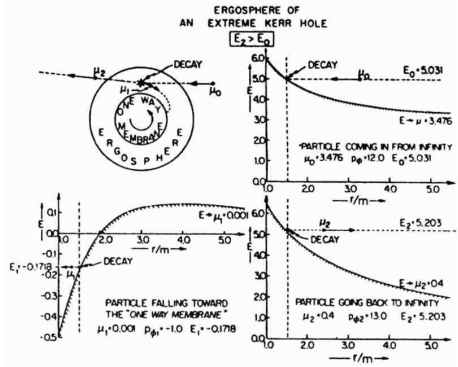}}
\end{center}
\vspace{-0.2cm}
\caption[]{(Reproduced from Ruffini and Wheeler with their kind permission.) Decay of a particle of rest-plus-kinetic energy $E_\circ$
     into a particle which is captured by the black hole with positive
     energy as judged locally, but negative energy $E_1$ as judged from infinity, together with a particle of rest-plus-kinetic energy $E_2>E_\circ$ which escapes to infinity. The cross-hatched curves give the effective potential (gravitational plus centrifugal) defined by the solution $E$ of Eq.(2) for constant values of $p_\phi$ and $\mu$ (figure and caption
reproduced from Christodoulou 1970 \cite{chris1}).}
\label{pic1}
\end{figure}
The first specific example of such an energy extraction process by a gedanken experiment was given using the above mentioned effective potential technique in Ruffini and Wheeler (1970) \cite{ruffx}, see Figure (\ref{pic1}),
and later by Floyd and Penrose (1971)\cite{fr}.
The reason for showing this figure here is a) to recall the first explicit computation and b) the introduction of the ``ergosphere", the region between the horizon of a Kerr-Newmann metric and the surface of infinite redshift were the enervy extraction process can occur, and also c) to emphasize how contrived, difficult but also conceptually novel such a mechanism of energy extraction can be. It is a phenomenon which is not localize at a point but can occur in an entire region: a global effect which relies essentially on the concept of a field. It can only work, however, for very special parameters and is in general associated with a reduction of the rest mass of the particle involved in the process. To slow down the rotation of a black hole and to increase its horizon by  accretion of counter-rotating particles is almost trivial, but to extract the rotational energy from a black hole by a slowing down process is extremely difficult, as also clearly pointed out by the example in Figure (\ref{pic1}).

The establishment of this analogy offered us the opportunity to appreciate once more the profound difference of seemingly similar effects in General relativity and classical field theories. In addition to the existence of totally new phenomena, just to mention one let us recall the dragging of the inertial frames around a rotating Black Hole,we had the first glimpse to an entire new field of theoretical physics present in and implied by the field equations of general relativity. The deep discussions on these problematic with Demetrios Christodoulou,, who was at the time in Princeton at the age of 17,  my first graduate student, lead us to the discovery of  the existence in Black Holes Physics both of "reversible and irreversible transformations"

\section{The first Analogy between Thermodynamics and  Black Hole physics : the reversible and irreversible transformations}

\begin{figure}
\vspace{-.5cm}
\epsfxsize=6.0cm
\begin{center}
\mbox{\epsfbox{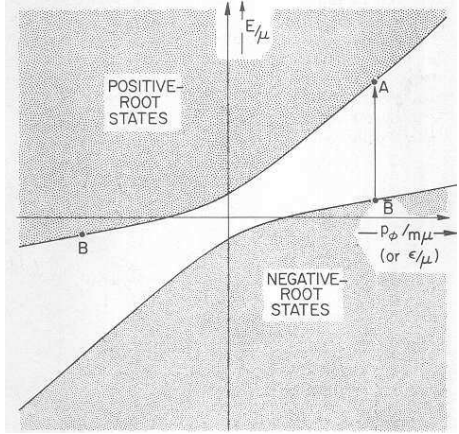}}
\end{center}
\vspace{-0.2cm}
\caption[]{Reversing the effect of having added to the black hole one particle (A) by adding another particle (B) of the same rest mass but opposite angular momentum and charge in a ``positive-root negative-energy state''. Addition of B is equivalent to subtraction of $B^-$. Thus the combined effect of the capture of particles A and B is an increase in the mass of the black hole given by the vector $B^-A$. This vector vanishes and reversibility is achieved when and only when the separation between positive root states and negative root states is zero, in this case the hyperbolas coalesce to a straight line. Reproduced from \cite{ruffc}.}
\label{posneg}
\end{figure}

It was indeed by analyzing the capture of test particles by an EMBH that we identified a set of limiting transformations which did not affect the surface area of an EMBH. These special transformations had to be performed very slowly, with a limiting value of zero kinetic energy on the horizon of the EMBH, see Fig.~\ref{posneg}. It became then immediately clear that the total energy of an EMBH could in principle be expressed in a ``rest energy'' a ``Coulomb energy'' and a ``rotational energy''. The rest energy being ``irreducible, the other two being submitted to positive and negative variations, corresponding respectively to process of addition and extraction of energy.

Why Wheeler  was mainly addressing the issue of the thermodynamical analogy I address with Demetrios the fundamental issue of the energetics of EMBH using the tools of the reversible and irreversible transformations. 
We finally obtained the general mass-energy formula for black holes (Christodoulou and Ruffini 1971) \cite{ruffc}: 
\begin{eqnarray}
E^2&=&M^2c^4=\left(M_{\rm ir}c^2 + {Q^2\over2\rho_+}\right)^2+{L^2c^2\over \rho_+^2},\label{em}\\
S&=& 4\pi \rho_+^2=4\pi(r_+^2+{L^2\over c^2M^2})=16\pi\left({G^2\over c^4}\right) M^2_{\rm ir},
\label{sa}
\end{eqnarray}
with
\begin{equation}
{1\over \rho_+^4}\left({G^2\over c^8}\right)\left( Q^4+4L^2c^2\right)\leq 1,
\label{s1}
\end{equation}
where $M_{\rm ir}$ is the irreducible mass, $r_{+}$ is the horizon radius, $\rho_+$ is the quasi-spheroidal cylindrical coordinate of the horizon evaluated at the equatorial plane,
$S$ is the horizon surface area, and extreme black holes satisfy the equality in eq.~(\ref{s1}). The crucial point is that transformations at constant surface area of the black hole, namely reversible transformations, can release an energy up to 29\% of the mass-energy of an extremal rotating black hole and up to 50\% of the mass-energy of an extremely magnetized and charged black hole. Since my Les Houches lectures ``On the energetics of black holes" (B.C. De Witt 1973) \cite{dw73}, one of my main research goals has been to identify an astrophysical setting where the extractable mass-energy of the black hole could manifest itself. As we will see in the following, I propose that this extractable energy of an EMBH be the energy source of Gamma-Ray Bursts (GRBs).

The thermodynamical analogy was further developed by Wheeler. By this time I had become convinced that the establishment of  a one to one analogy between General Relativistic effects and classical results was totally hopeless. It appeared to me that the use of analogy could only be used as a  very important tools to explore, possibly rationalizable in a formula, the extremely vast theoretical world of the space-time structures contained in Einstein theory of general relativity. A good example is our mass formula. Trying to enforce a perfect analogy is too risky and reductive.

\section{The paradigm for the identification of the first "Black Hole"in our Galaxy and the development of X ray Astronomy.}

The launch of the ``Uhuru'' satellite dedicated to the first systematic examination of the Universe in X rays, by the group directed by Riccardo Giacconi at American Science and Engineering, signed a fundamental progress and generated a tremendous momentum in the field of Relativistic Astrophysics.  The very fortunate collaboration soon established with simultaneous observations in the optical and in the radio wavelengths allowed to have high quality data on binary star systems composed of a normal star being stripped of matter by a compact massive companion star: either a neutron star or a Black Hole.

The ``maximum mass of a neutron star" was the subject of the thesis
of Clifford Rhoades, my second graduate student at Princeton. A criteria was found there to overcome fundamental unknowns about the behaviour of matter at supranuclear densities by establishing an absolute upper limit to the neutron star mass based only on general relativity, causality and the behaviour of matter at nuclear and subnuclear densities.   This work, presented at the 1972 Les Houches summer School (B. and C. de Witt 1973), only appeared after a prolongued debate (see the reception and publication dates!) (Rhoades and Ruffini 1974) \cite{rr74}.

\begin{itemize}

\item 
The ``black hole uniqueness theorem", implying the axial symmetry of the
configuration and the absence of regular pulsations from black holes,
\item 
 the ``effective potential technique", determining the efficiency of the energy emission in the accretion process, and 
\item 
the ``upper limit on the maximum mass of a neutron star" discriminating between 
an unmagnetized neutron star and a black hole
\end{itemize}
were the three essential components in establishing the paradigm
for the identification of the first black hole in Cygnus X1 (Leach and Ruffini 1973) \cite{lr73}.
These results were also presented in a widely attended session chaired 
by John Wheeler at the 1972 Texas Symposium in New York, extensively reported by the New York Times. The New York Academy of Sciences which hosted the symposium had just awarded me their prestigious Cressy Morrison Award for my work on neutron stars and black holes. Much to their dismay I never wrote the paper for the proceedings since it coincided with the one submitted for publication (Leach and Ruffini 1973) \cite{lr73}. 

The definition of the paradigm did not
come easily but slowly matured after innumerable discussions, mainly on
the phone, both with Riccardo Giacconi and Herb Gursky. I still remember an irate
professor of the Physics Department at Princeton pointing publicly to my outrageous phone
bill of \$274 for one month, at the time considered scandalous, due to my
frequent calls to the Smithsonian, and a much more
relaxed and sympathetic attitude about this situation by the department chairman, Murph Goldberger. 
This work was finally summarized in two books: one with Herbert Gursky (Gursky and Ruffini 1975) \cite{gr75}, following the 1973 AAAS Annual Meeting in San Francisco, and the second with Riccardo Giacconi (Giacconi and Ruffini 1978) \cite{gr78} following the 1975 LXV Enrico Fermi Summer School (see also the proceedings of the 1973 Solvay Conference).

\section{More analogies between  thermodynamics and Black Holes physics}

The analogy between Thermodynamics and General relativity became in 1971 the topic of the specific Ph. D. Thesis of Wheeler's student Jacob Bekenstein. Through a profound set of gedanken experiments, Jacob pushed further the analogy between thermodynamics and Black Holes physics. Demetrios and I had formally established the existence of reversible transformations in Black Hole physics as well as the monotonic increase, as occurs for entropy in thermodynamics, of the irreducible mass $M_{\rm irr}$ of a Black Hole (from which the word ``irreducible'' arises) also formally established independently by Hawking in his area theorem \cite{h71}. The complete equivalence between the two results immediately follows from the identity relating the surface area $S = 16\pi M_{\rm irr}^2$ of the Black Hole to $M_{\rm irr}$, as suggested by Bryce Dewitt and confirmed in a quick calculation by Demetrios and myself. Jacob went one step further proposing that the area of the Black Hole S measured in Planck-Wheeler units should indeed be identified with entropy \cite{b73}. He did this by formulating a statistical interpretation of Black Hole entropy and introducing the first generalized form of the first law of thermodynamics in physical processes involving Black Holes. These topics even today, thirty years later, still inspire lively debate! Jacob's proposal was extremely interesting and very intriguing at the time and remains so for me in some ways even today. The proposal certainly was not contradictory but I could not find a necessity for transforming it into an identity. This entire matter became the subject of even more lively discussion after Stephen Hawking proposed a physical process which, if true (in the sense used by Wigner), would transform all these theoretical conjectures into physical reality: the Black Hole quantum evaporation process \cite{h74}. This topic also inspires lively debate some three decades later. It is likely that these issues will be clarified once there is a theory encompassing both General Relativity and Relativistic Quantum Field Theories.
	The basic formulae describing the Hawking radiation process from a Black Hole can be simply summarized in the following three formulae:
\begin{eqnarray}
{\rm Radiation\, temperature} &\,& T \simeq 0.62 \cdot 10^{-7}(M_{\odot} / M)K\nonumber\\
{\rm Evaporation\, time} &\,&\tau \simeq E/(dE/dt)\sim 2\cdot 10^{63}(M / M_{\odot})^3 years\\
{\rm Energy\, flux} &\,& dE/dt \simeq 10^{-22} (M_{\odot} /M )^2 erg/sec \nonumber
\end{eqnarray}

\section{Analogy between the electrodynamics of a Black Hole and a perfect conducting sphere }

\begin{figure}
\vspace{-.5cm}
\epsfxsize=6.0cm
\begin{center}
\mbox{\epsfbox{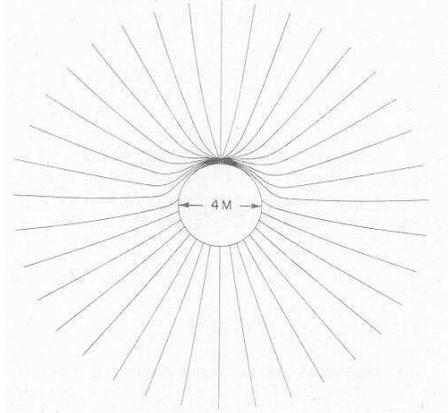}}
\end{center}
\vspace{-0.2cm}
\caption[]{Line of force of a charge near a black hole. Reproduced from \cite{hr73}.}
\label{linef}
\end{figure}

\begin{figure}
\vspace{-.5cm}
\epsfxsize=6.0cm
\begin{center}
\mbox{\epsfbox{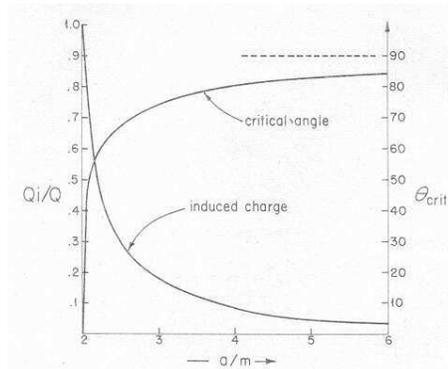}}
\end{center}
\vspace{-0.2cm}
\caption[]{Induced charge on a Schwarzschild black hole. Reproduced from \cite{hr73}.}
\label{indcharge}
\end{figure}

Before closing this set of analogies I would like to recall a gedanken experiment we described by the integration of the general relativistic equations with Richard Hanni back in 1974. The idea was to compute and draw the lines of force of a test charge in the field of a Schwarzschild Black Hole and was the topic of the senior thesis of Rick, then my undergraduate student at Princeton. In order to solve this problem we decided to introduce the concept of an ``induced charge'' on a Black Hole \cite{hr73}. By doing so we proposed the analogy between a Black Hole and a classical system with a surface equal to the horizon and endowed of an appropriate conductivity. See Fig.~\ref{linef}  and Fig.~\ref{indcharge}. This gedanken experiment opened the way to the field of Black Holes electrodynamic which was further expanded in important contributions  by Damour, Hanni, Ruffini and Wilson \cite{dhrw78} and references therein and by Thibault Damour. Thibault in 1974 came to Princeton to work with me on his state doctorate thesis to be discussed in Paris at the Ecole Normale Superieure. The subject of the beautiful thesis of Thibault incorporated a detailed discussion of the general relativistic effects of electrodynamics of EMBH including generalized Ohm's, Joule's, Ampere's , Navier Stokes Laws using, as a tool, the above mentioned analogy. 
Interestingly this analogy was taken very seriously and written up in a book by Kip Thorne and collaborators \cite{membrane} as a final theory of Black Holes. See however subtle relativistic effects and different conclusions  reached by Brian Punsly in his recent interesting book \cite{punsly}.

This analogy was reconsidered in a widely publicized article in a full page on the Frankfurther Allgemeine, shown to me by Hagen a few months ago, referring to an article by Wilczek and a collaborator \cite{Wilczek}. In that paper Wilczek, without giving any reference, purports to extend our results with Hanni on the induced charge on a Schwarzschild Black Hole to the case of a Reissner Nordstr\"{o}m geometry with charge Q obtaining the very simple result for the effective potential derived from the image charge technique q: 
$V(r)=q\left( \frac{Q-q}{r}-\frac{qR}{r^2-R^2}\right)$

\begin{figure}
\vspace{-.5cm}
\epsfxsize=6.0cm
\begin{center}
\mbox{\epsfbox{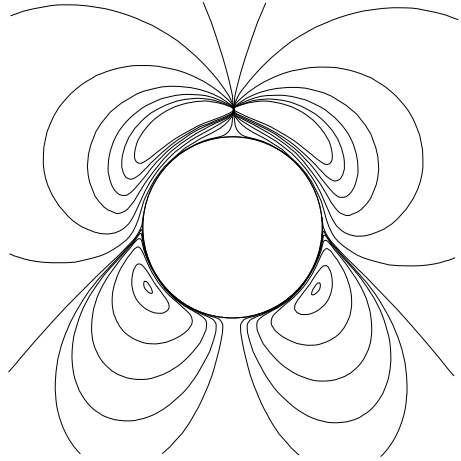}}
\end{center}
\vspace{-0.2cm}
\caption[]{Line of force of a point charge near an extreme EMBH with $Q=M$. Reproduced from \cite{bicak}.}
\label{linefbic}
\end{figure}

However in a very interesting set of papers Bicak and his student Dvorak \cite{bicak} have proven a totally new situation in the lines of force of a charge in the presence of a Reissner Nordtr\"{o}m geometry, see Fig.~\ref{linefbic}. In other words, in the limit of Q/M=1, an extreme Black Hole, no line of force from the test particle crosses the horizon. We are in presence of a totally new effect for the electric field due to General Relativity reminiscent of the Meissner effect in classical magnetic fields in presence of superconductors. Such an electric general relativistic  Meissner effect does not appear to be contemplated in the Wilczeck solution given above developed by analogy to my solution with Hanni. This case can be considered a propedeutic example of the incomparable richness of physical regimes present in the General Relativity which, once again, do transcend the direct analogies with classical field theories.

This series of events offer a clear pedagogical example of how the enforcement of direct and unproven analogies in General Relativity can be at times dangerous and lead to incorrect conclusions.

\section{The discovery of Gamma Ray Bursts}

\begin{figure}
\vspace{-.5cm}
\epsfxsize=6.0cm
\begin{center}
\mbox{\epsfbox{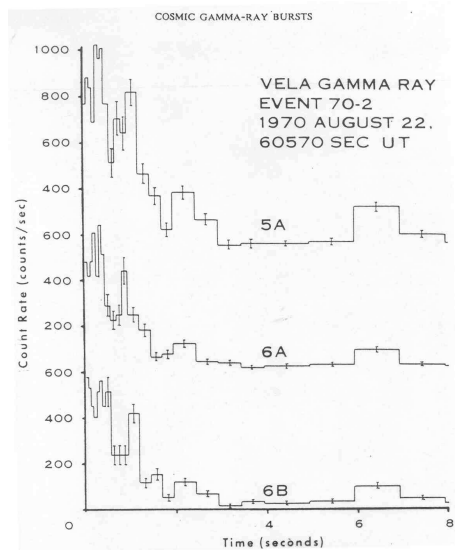}}
\end{center}
\vspace{-0.2cm}
\caption[]{One of the first GRBs observed by the Vela satellite. Reproduced from Strong in Gursky \& Ruffini (1975) \cite{gr75}.}
\label{velaburst}
\end{figure}

\begin{figure}
\vspace{-.5cm}
\epsfxsize=\hsize
\begin{center}
\mbox{\epsfbox{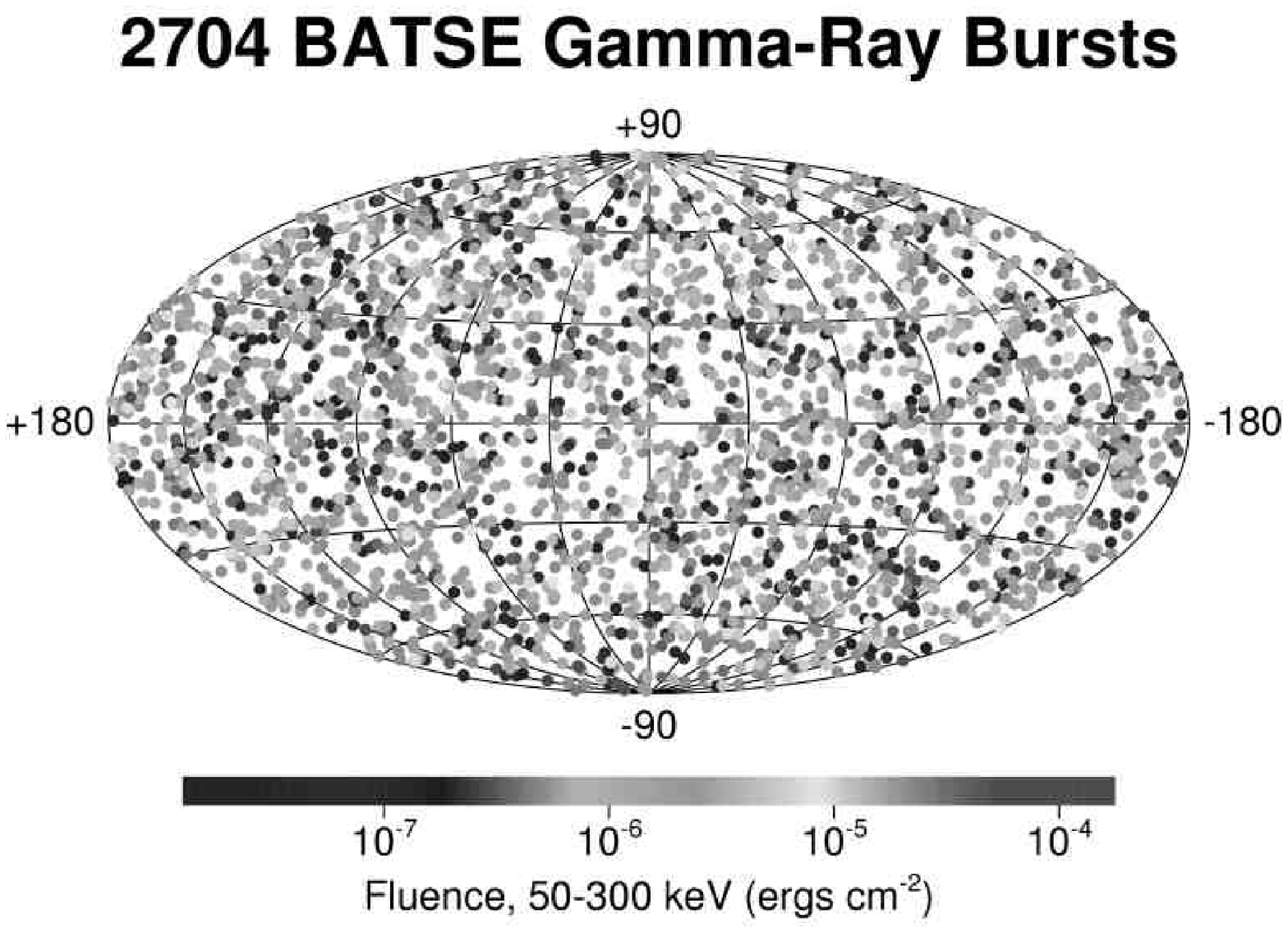}}
\end{center}
\vspace{-0.2cm}
\caption[]{Angular distribution of GRBs in galactic coordinates from the Compton GRO satellite.}
\label{batsedist}
\end{figure}

In 1975 Herbert Gursky and myself had been invited by the AAAS to organize for their annual meeting in S. Francisco a session on Neutron Stars, Black Holes and Binary X ray sources. During the preparation of the meeting we heard that some observations made from the military Vela satellites, conceived in order to monitor the Limited Test Ban Treaty of 1963 on atomic bomb explosion, had just been unclassified and we asked Ian B. Strong to report for the first time in a public meeting on them: the gamma ray bursts (Strong 1975) \cite{gr75}. See Fig.~\ref{velaburst}. It was clear since the earliest observations that these signals were not coming either from the Earth nor the planetary system. By 1991 a great improvement on the distribution of the GRB came with the launch by NASA of the Compton Gamma-Ray Observatory which in ten years of observations gave a beautiful evidence for the perfect isotropy of the angular distribution of the GRB source in the sky, see Fig.~\ref{batsedist}. The sources had to be either at cosmological distances or very close to the solar system so as not to feel the galactic anisotropical distribution. In the mean time the number of theories grew exponentially but without any clear conclusion. Quite prominent among these theories were some relating the GRB phenomenon to the Hawking radiation process.

\section{Analogy of the Heisenberg-Euler critical capacitor and the vacuum polarization around a macroscopic Black Hole}

\begin{figure}
\vspace{-.5cm}
\epsfxsize=\hsize
\begin{center}
\mbox{\epsfbox{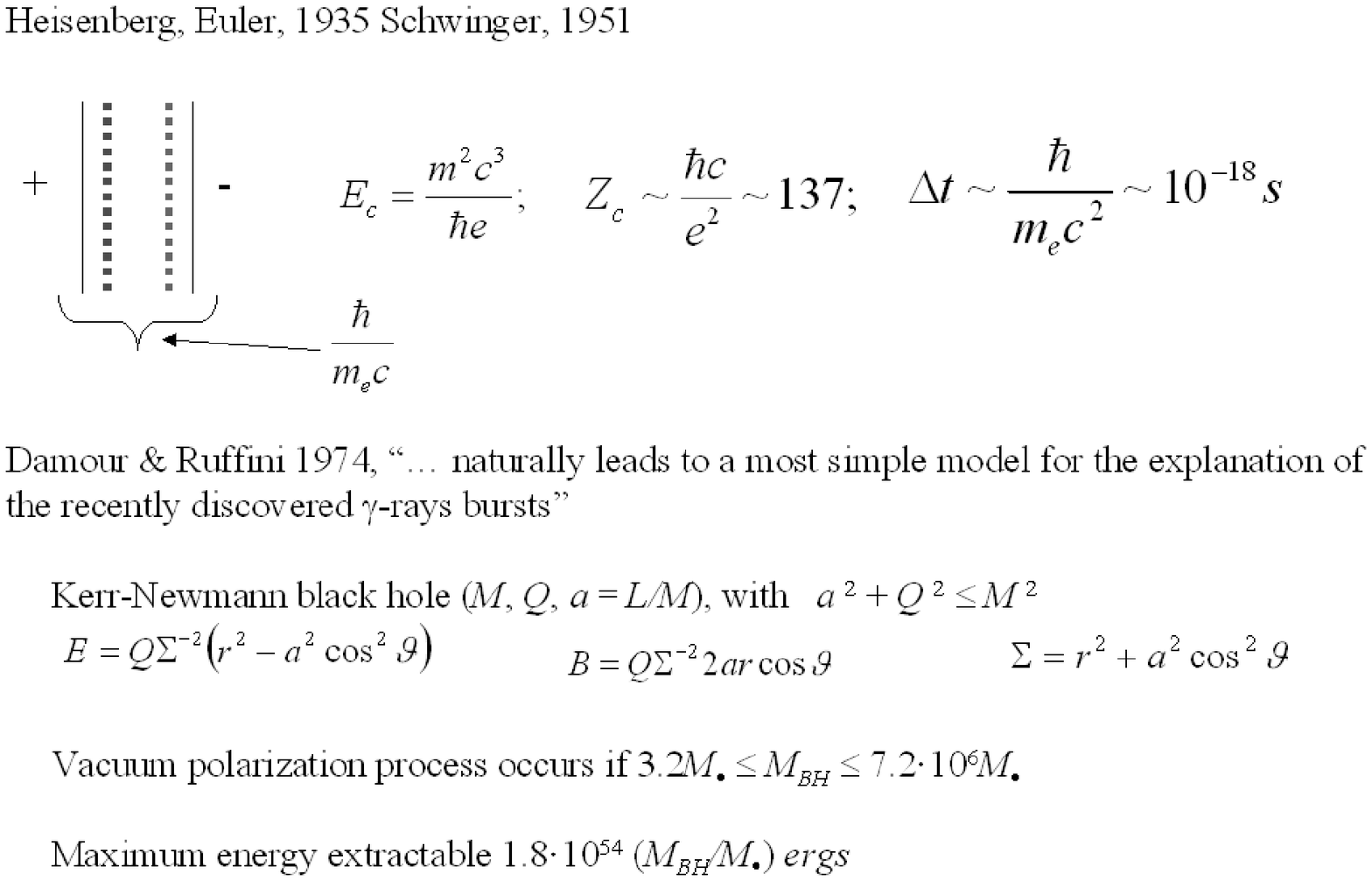}}
\end{center}
\vspace{-0.2cm}
\caption[]{Summary of the EMBH vacuum polarization process. See Damour \& Ruffini (1975) \cite{dr75} for details.}
\label{capdiap}
\end{figure}

In 1975, following the work on the energetics of black holes (Christodoulou and Ruffini 1971) \cite{ruffc}, we pointed out (Damour and Ruffini, 1975) \cite{dr75} the existence of the vacuum polarization process {\it a' la} Heisenberg-Euler-Schwinger (Heisenberg and Euler 1935 \cite{he35}, Schwinger 1951 \cite{s51}) around black holes endowed with electromagnetic structure (EMBHs). Such a process can only occur for EMBHs of mass smaller then $7.2\cdot 10^{6}M_\odot$. The basic energetics implications were contained in Table~1 of that paper (Damour and Ruffini, 1975) \cite{dr75}, where it was also shown that this process is almost reversible in the sense introduced by Christodoulou and Ruffini (1971) \cite{ruffc} and that it extract the mass energy of an EMBH very efficiently. We also pointed out that this vacuum polarization process around an EMBH offered a natural mechanism for explaining GRBs and the characteristic energetics of the burst could be $\ge 10^{54}$ ergs, see Fig.~\ref{capdiap}.

\section{The Beppo-SAX Satellite: the instantaneous destruction of more then 135 theoretical models}

It was only with the very unexpected and fortuitous observations of the Beppo-SAX satellite that the existence of a long lasting afterglow of these sources was identified: this has led to the determination of a much more accurate position for these sources in the sky, which permitted in turn, for the first time, their optical and radio identification. The optical identification has led to the determination of their cosmological distances and to their paramount energetic requirements in some cases $\ge 10^{54}$ ergs (see Costa 2001 \cite{c01}).

The very fortunate interaction and resonance between X-ray, optical and radio astronomy which in the seventies allowed the maturing of the physics and astrophysics of neutron stars and black holes (see e.g. Giacconi and Ruffini 1978 \cite{gr78}) promises to be active again today in unravelling the physics and astrophysics of the gamma ray burst sources.

\begin{figure}
\vspace{-.5cm}
\epsfysize=17.0cm
\begin{center}
\mbox{\epsfbox{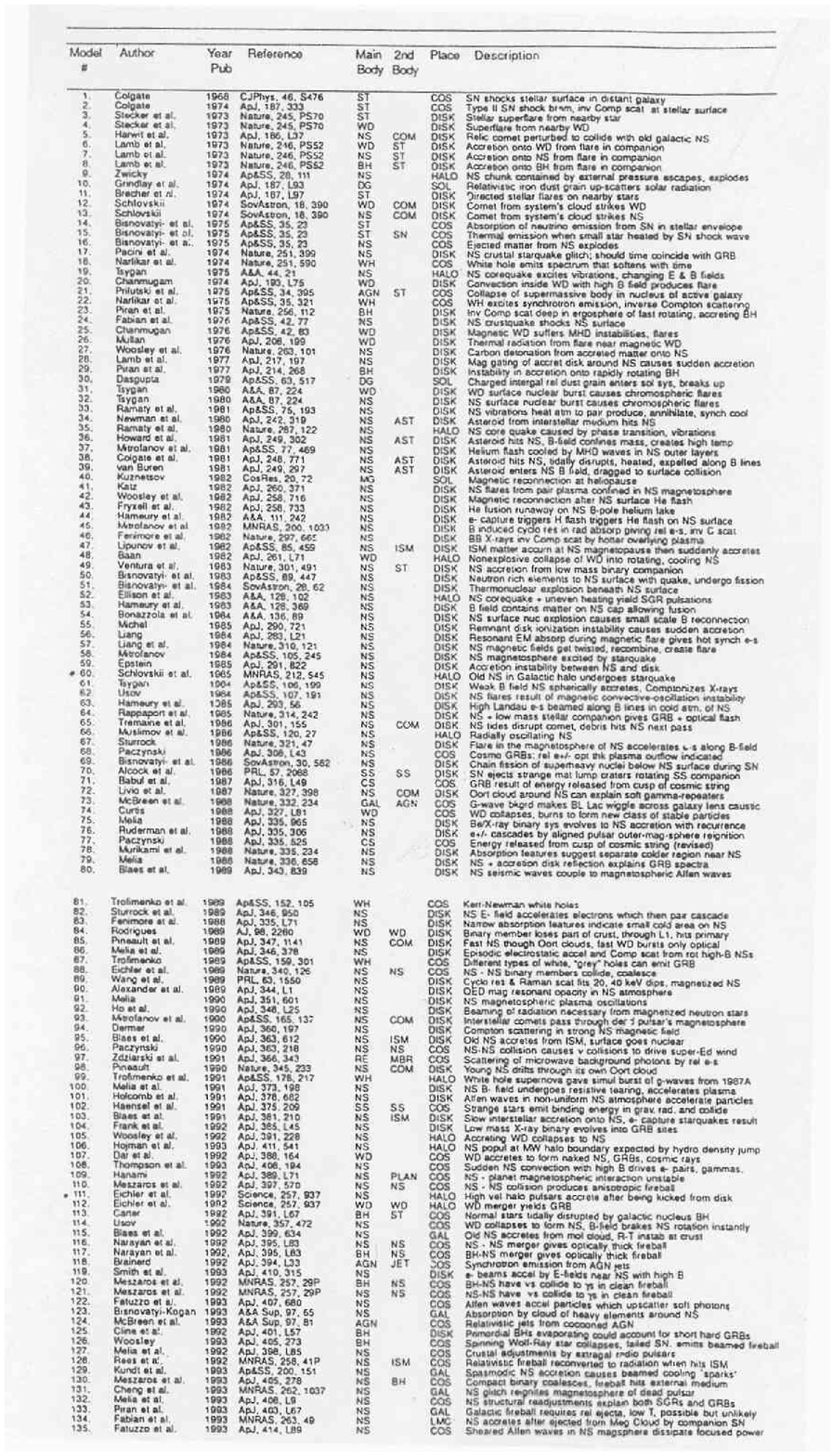}}
\end{center}
\vspace{-0.2cm}
\caption[]{Partial list of theories before the Beppo-SAX, from a talk presented at MGIXMM \cite{mgixmm}.}
\label{teofig}
\end{figure}

The observations of the Beppo-SAX satellite had a very sobering effect on the theoretical developments on GRB models. Practically the totality of the existing theories, see a partial list in Fig.~\ref{teofig}, were at once wiped out, not being able to fit the stringent energetics requirements imposed by the observations. Particularly constraining the observations were for the models based on the Hawking radiation process. See Tab.~\ref{energ} for details.

\begin{table}
\caption[]{Hawking radiation process vs. GRB observations.}
\footnotesize
\vspace{0.5cm}
For the correct value of the energetics $E_{tot} \simeq 10M_{\odot} \simeq 10^{55}erg$, we have:\\
\begin{center}
\begin{tabular}{|c|c|c|}
\hline
Theoretical value & Observed value & Discrepancy \\
\hline
 & & \\
$T=6.2\cdot10^{-9}$K  &  $T \simeq 10^8$K  & $ \sim 10^{-17}$  \\
 & & \\
$\tau=10^{73}$sec  &  $\tau \simeq 1$sec & $\sim 10^{73}$ \\
 & & \\
$\left(\frac{dE}{dt}\right)=10^{-24} erg/sec$  &  $\left(\frac{dE}{dt}\right) \simeq 10^{54} erg/sec$  &  $\sim 10^{-78}$\\
 & & \\
\hline
\end{tabular}
\end{center}
\vspace{0.5cm}
For the correct value of the time scale $\tau \simeq 1$ sec, we have:\\
\begin{center}
\begin{tabular}{|c|c|c|}
\hline
Theoretical value & Observed value & Discrepancy \\
\hline
 & & \\
$E_{tot} \simeq 10^{-24} M_{\odot} \simeq 10^{30}erg$  &  $E_{tot} \simeq 10^{55}$erg & $\sim 10^{-25}$ \\
 & & \\
$T=10^{17}$K  &  $T \simeq 10^{8}$K & $\sim 10^{9}$ \\
 & & \\
$\left(\frac{dE}{dt}\right)=10^{26} erg/sec$  &  $\left(\frac{dE}{dt}\right) \simeq 10^{54} erg/sec$  &  $\sim 10^{-28}$\\
 & & \\
\hline
\end{tabular}
\end{center}
\vspace{0.5cm}
For the correct value of the spectrum energy $T=10^{8}$K, we have:\\
\begin{center}
\begin{tabular}{|c|c|c|}
\hline
Theoretical value & Observed value & Discrepancy \\
\hline
 & & \\
$E_{tot} \simeq 10^{-9} M_{\odot} \simeq 10^{45}erg$  &  $E_{tot} \simeq 10^{55}$erg & $\sim 10^{-10}$ \\
 & & \\
$\tau=10^{43}$sec  &  $\tau \simeq 1$sec & $\sim 10^{43}$ \\
 & & \\
$\left(\frac{dE}{dt}\right)=10^{-4} erg/sec$  &  $\left(\frac{dE}{dt}\right) \simeq 10^{54} erg/sec$  &  $\sim 10^{-58}$\\
 & & \\
\hline
\end{tabular} 
\end{center}  
\label{energ}    
\end{table}

This is by far the theoretical predictions furthest from any observational data in the entire history of the Homo Sapiens and possibly in the entire Universe.

\section{Analogy between the Ergosphere and the  Dyadosphere of a Black Hole}

The enormous energy requirements of GRBs, very similar to the ones predicted in Damour \& Ruffini (1975) \cite{dr75} have convinced us to return to our earlier work in studying more accurately the process of vacuum polarization and the region of pair creation around an EMBH. This has led a) to the new concept of the dyadosphere of an EMBH (named for the Greek word for pair) and b) to the concept of a plasma-electromagnetic (PEM) pulse and c) to the analysis its temporal evolution generating signals with the characteristic features of a GRB.

In our theoretical approach, we imply that by the observations of GRBs we are witnessing the formation of an EMBH and we so follow the process of gravitational collapse in real time. Even more important, the tremendous energies involved in the energetics of these sources have their origin in the extractable energy of black holes given in Eqs.~(1)--(3) above.

Various models have been proposed in order to extract the rotational energy of black holes by processes of relativistic magnetohydrodynamics (see e.g., Ruffini and Wilson (1975) \cite{rw75}). It should be expected, however, that these processes are relevant over the long time scales characteristic of accretion processes.

In the present case of gamma ray bursts a sudden mechanism appears to be at work on time scales of the order of few seconds or shorter and they are naturally explained by the vacuum polarization process introduced in Damour \& Ruffini (1975) \cite{dr75}.

The fundamental new points we have found re-examining our previous work can be simply summarized, see Preparata, Ruffini and Xue (1998a,b) \cite{prx98ab} for details:

\begin{itemize}

\item 
The vacuum polarization process can  occur in an extended region around the black hole called the dyadosphere, extending from the horizon radius $r_+$ out to the dyadosphere radius $r_{ds}$. Only black holes with a mass larger than the upper limit of a neutron star and up to a maximum mass of $7.2\cdot 10^{6}M_\odot$ can have a dyadosphere.

\item 
The efficiency of transforming the mass-energy of a black hole into particle-antiparticle pairs outside the horizon can approach 100\%, for black holes in the above mass range. 

\item 
The pair created are mainly positron-electron pairs and their number is much larger than the quantity $Q/e$ one would have naively expected on the grounds of qualitative considerations. It is actually given by $N_{\rm pairs}\sim{Q\over e}{r_{ds}\over \hbar/mc}$, where $m$ and $e$ are respectively  the electron mass and charge.  The energy of the pairs and consequently the  emission of the associated electromagnetic radiation peaks in the gamma X-ray region, as a function of the black hole mass.

\end{itemize}

Let us now recall the main results on the dyadosphere obtained in Preparata, Ruffini and Xue (1998a,b) \cite{prx98ab}. Although the general considerations presented by Damour and Ruffini (1975) \cite{dr75} did refer to a Kerr-Newmann field with axial symmetry about the rotation axis, for simplicity, we have there considered the case of a nonrotating Reissner-Nordstr\"{o}m EMBH to illustrate the basic gravitational-electrodynamical process. The dyadosphere then lies between the radius 
\begin{equation}
r_{\rm ds}=\left({\hbar\over mc}\right)^{1\over2}
\left({GM\over c^2}\right)^{1\over2} 
\left({m_{\rm p}\over m}\right)^{1\over2}
\left({e\over q_{\rm p}}\right)^{1\over2}
\left({Q\over\sqrt{G} M}\right)^{1\over2}.
\label{rc}
\end{equation} 
and the horizon radius 
\begin{equation}
r_{+}={GM\over c^2}\left[1+\sqrt{1-{Q^2\over GM^2}}\right].
\label{r+}
\end{equation}
The number density of pairs created in the dyadosphere is 
\begin{equation}
N_{e^+e^-}\simeq {Q-Q_c\over e}\left[1+{
(r_{ds}-r_+)\over {\hbar\over mc}}\right] \ ,
\label{n}
\end{equation}
where $Q_c=4\pi r_+^2{m^2c^3\over \hbar e}$. The total energy of pairs, converted from the static electric energy,
deposited within the dyadosphere is then
\begin{equation}
E^{\rm tot}_{e^+e^-}={1\over2}{Q^2\over r_+}(1-{r_+\over r_{\rm ds}})(1-
\left({r_+\over r_{\rm ds}}\right)^2) ~.
\label{tee}
\end{equation}

The analogies between the ergosphere and the dyadosphere are many and extremely attractive:

\begin{itemize}
\item Both of them are extended regions around the black hole.
\item In both regions the energy of the black hole can be extracted, approaching the limiting case of reversibility as from Christodoulou \& Ruffini (1971) \cite{ruffc}.
\item The electromagnetic energy extraction by the pair creation process in the dyadosphere is much simpler and less contrived than the corresponding process of rotational energy extraction from the ergosphere.
\end{itemize}

\section{Analogies between an EM pulse of an atomic explosion and the PEM pulse of a Black Hole}

The analysis of the radially resolved evolution of the energy
deposited within the $e^+e^-$-pair and photon plasma fluid created
in the dyadosphere of an EMBH is much more complex then we had initially anticipated. The collaboration with Jim Wilson and his group at Livermore Radiation Laboratory has been very important to us. We decided to join forces and propose a new collaboration with the Livermore group renewing the successful collaboration with Jim of 1974 (Ruffini and Wilson 1975) \cite{rw75}. We proceeded in parallel: in Rome with simple almost analytic models to be then validated by the Livermore codes (Wilson, Salmonson and Mathews 1997,1998) \cite{wsm97} \cite{wsm98}.

For the evolution we assumed the relativistic hydrodynamic equations, for details see Ruffini, et al. (1998,1999) \cite{rswx98} \cite{rswx99}.
We assumed the plasma fluid of $e^+e^-$-pairs, photons and baryons to be a simple perfect fluid in the curved space-time. The baryon-number and energy-momentum conservation laws are 
\begin{eqnarray}
(n_B U^\mu)_{;\mu}&=&(n_BU^t)_{,t}+{1\over r^2}(r^2 n_BU^r)_{,r}= 0~,
\label{contin}\\
(T_\mu^\sigma)_{;\sigma}&=&0,
\label{contine}
\end{eqnarray}
and the rate equation: 
\begin{equation}
(n_{e^\pm}U^\mu)_{;\mu}=\overline{\sigma v} \left[n_{e^-}(T)n_{e^+}(T) - n_{e^-}n_{e^+}\right] ~,
\label{econtin}
\end{equation}
where $U^\mu$ is the four-velocity of the plasma fluid, $n_B$ the proper baryon-number density, $n_{e^\pm}$ are the proper densities of electrons and positrons($e^\pm$), $\sigma$ is the mean pair annihilation-creation cross-section, $v$ is the thermal velocity of $e^\pm$, and $n_{e^\pm}(T)$ are the proper number-densities of $e^\pm$ at an appropriate equilibrium temperature $T$. The calculations are continued until the plasma fluid
expands, cools and the $e^+e^-$ pairs recombine and the system becomes optically thin.

The results of the Livermore computer code are compared and contrasted with three almost analytical models: 
(i) spherical model: the radial component of
four-velocity is of the form $U(r)=U{r\over {\cal R}}$, where $U$ is the four-velocity at
the surface (${\cal R}$) of the plasma, similar to a portion of a Friedmann model 
(ii) slab 1: $U(r)=U_r={\rm
const.}$, an expanding slab  with constant width ${\cal D}= R_\circ$ in
the coordinate frame in which the plasma is moving; 
(iii) slab 2: 
an expanding slab with constant width $R_2-R_1=R_\circ$ in the comoving frame of the plasma. 

We compute the relativistic Lorentz factor $\gamma$ of the expanding
$e^+e^-$ pair and photon plasma.

\begin{figure}
\vspace{-.5cm}
\epsfxsize=\hsize
\begin{center}
\mbox{\epsfbox{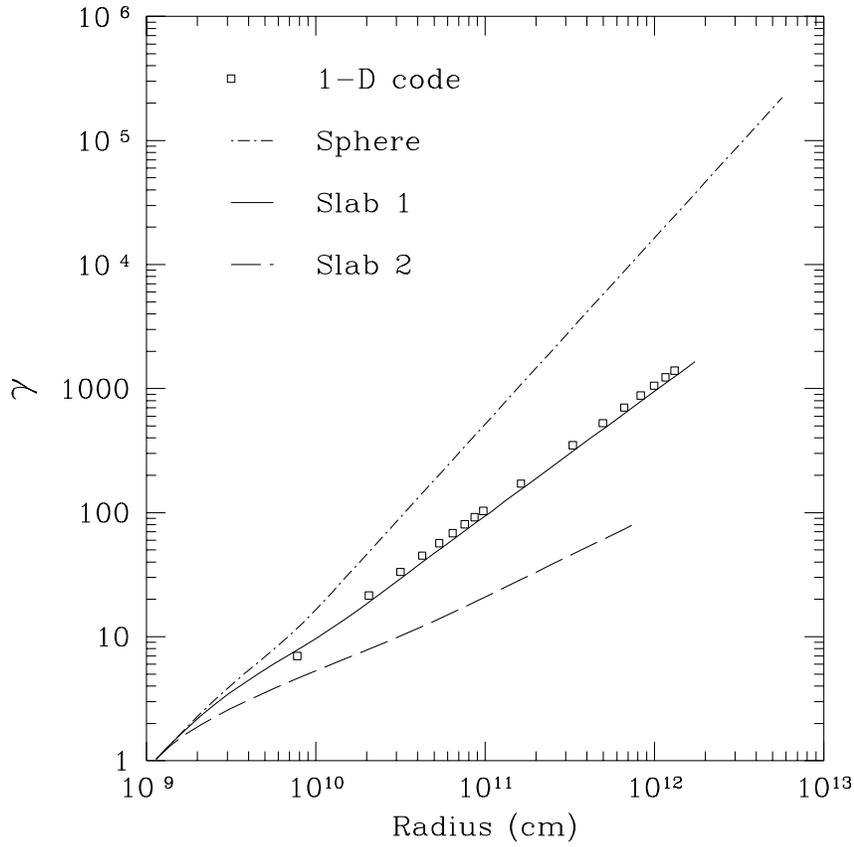}}
\end{center}
\vspace{-0.2cm}
\caption[]{Lorentz $\gamma$ as a function of radius.
Three models for the expansion pattern of the $e^+e^-$ pair plasma are
compared with the results of the one dimensional hydrodynamic code for
a $1000 M_\odot$ black hole with charge $Q = 0.1 Q_{max}$.  The 1-D
code has an expansion 
pattern that strongly resembles that of a shell
with constant coordinate thickness. Reproduced from Ruffini, et al. (1999) \cite{rswx99}.}
\label{pic2}
\end{figure}

In Figure (\ref{pic2}) we see a comparison of the
Lorentz factor of the expanding fluid as a function of radius for all
the models. We can see that the one-dimensional code (only a few 
significant points 
are plotted) matches the
expansion pattern of a shell of constant coordinate thickness.

In analogy with the notorious electromagnetic radiation EM  pulse  of some explosive events, we called this relativistic counterpart 
of an expanding pair electromagnetic radiation shell a PEM pulse.

\begin{figure}
\vspace{-.5cm}
\epsfxsize=\hsize
\begin{center}
\mbox{\epsfbox{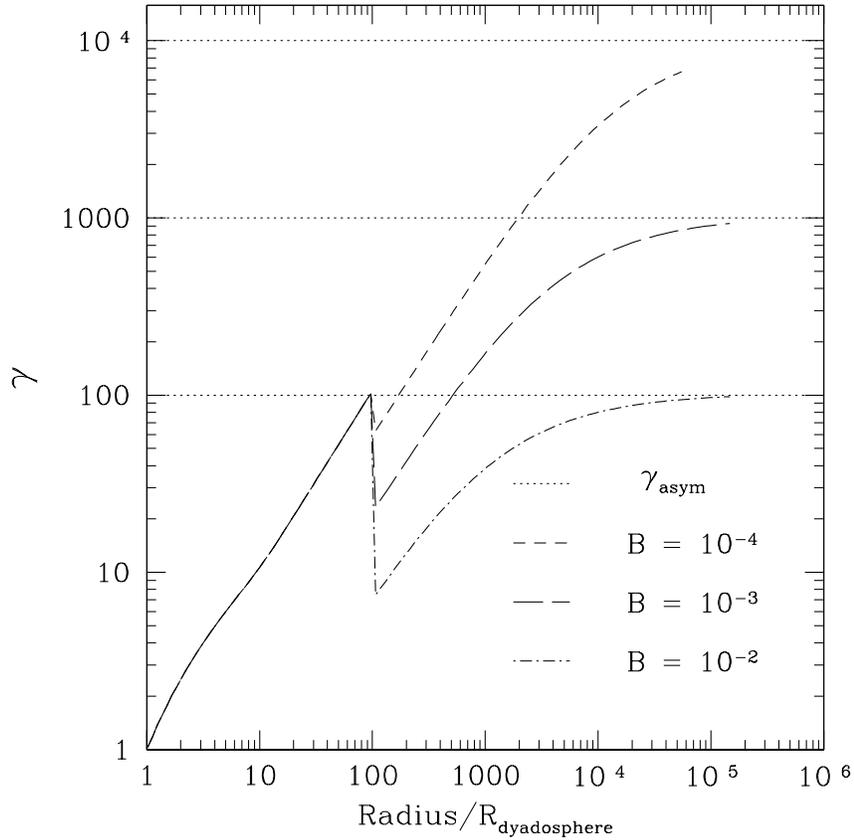}}
\end{center}
\vspace{-0.2cm}
\caption[]{Lorentz $\gamma$ factor as a function of radius for the PEM pulse interacting with the baryonic matter of the remnant (PEMB pulse) for selected values of the baryonic matter. Reproduced from Ruffini, et al. (2000) \cite{rswx00}.}
\label{gammab}
\end{figure}

\begin{figure}
\vspace{-.5cm}
\epsfxsize=\hsize
\begin{center}
\mbox{\epsfbox{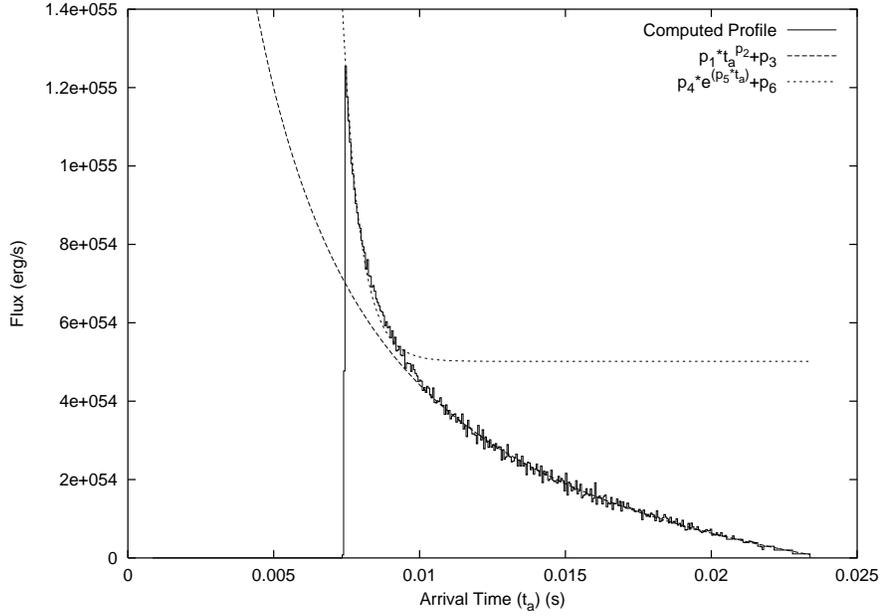}}
\end{center}
\vspace{-0.2cm}
\caption[]{P-GRB from an EMBH with $M=100 M_\odot$ and $Q = 0.1 Q_{max}$. Reproduced from Bianco, et al. (2001) \cite{brx01}.}
\label{pic3}
\end{figure}

In recent works we have computed the interaction of the expanding plasma with the surrounding baryonic matter (Ruffini, et al. 2000) \cite{rswx00}, see Fig~.\ref{gammab}. We have also been able to follow the expansion process all the way to the point where the transparency condition is reached and what we have defined the ``Proper GRB'' (P-GRB) is emitted (Bianco, et al. 2001) \cite{brx01}, see Fig~\ref{pic3}. These results of our theoretical model have reached the point to be submitted to a direct comparison with the observational data.

\section{The three new paradigms for the interpretation of GRBs}

Starting from this theoretical background, we have moved ahead to fit for the first time the observational data on the ground of EMBH model. We have used, as a prototype, the GRB~991216, both for its very high energetics, which we have estimated in the range of $E_{\rm dya}\sim 9.57\times 10^{52}$ ergs, and for the superb data obtained by the Chandra and RXTE satellites. We have found a necessity to formulate in our novel approach three new paradigms, in order to understand the GRB phenomenon:

\begin{enumerate}
\item The Relative Space-Time Transformations (RSTT) paradigm. See Ruffini, Bianco, Chardonnet, Fraschetti, Xue (2001a) \cite{lett1}.
\item The Interpretation of the Burst Structure (IBS) paradigm. See Ruffini, Bianco, Chardonnet, Fraschetti, Xue (2001b) \cite{lett2}.
\item The Multiple-Collapse Time Sequence (MCTS) paradigm. See Ruffini, Bianco, Chardonnet, Fraschetti, Xue (2001c) \cite{lett3}.
\end{enumerate}

These results are currently under refereeing process in ApJ Letters since 28/11/2000.

\section{conclusions}

From the above experience, I can venture to formulate some conclusions, which may be of general validity:

\subsection{On analogies}

\begin{itemize}
\item Analogies have been extremely helpful in establishing similarities and deepening our physical understanding if applied to two circumstances both derived within a general relativistic framework. The analogies between dyadosphere and ergosphere are good examples.
\item The analogies between classical regimes and general relativistic regimes have been at times helpful in giving the opportunity to glance on the enormous richness of the new physical processes contained in Einstein's theory of space-time structure. In some cases they have allowed to reach new knowledge and formalize new physical laws, the derivation on Eqs.(1)--(3) is a good example. Such analogies have also dramatically evidenced the enormous differences in depth and physical complexity between the classical physics and general relativistic effects. The case of extraction of rotational energy from a neutron star and a rotating black hole are a good example.
\item In no way an analogy based on classical physics can be enforced on general relativistic regimes. Such an analogy is too constraining and the relativistic theory shows systematically a wealth of novel physical circumstances and conceptual subtleties, unreachable within a classical theory. The analogies in the classical electrodynamics we just outlined are good examples.
\end{itemize}

\subsection{On new paradigms}

The establishment of new paradigms is essential to the scientific process, and certainly not easy to obtain. Such paradigms are important in order to guide a meaningful comparison between theories and observations and much attention should be given to their development and inner conceptual consistency.

\subsection{On observational data}

Always the major factors in driving the progress of scientific knowledge is the confrontation of the theoretical predictions with the observational data. The evolution of new technologies has allowed, in recent years, to dramatically improve the sensitivity of the observational apparata. It is very gratifying that, in this process of learning the structure of our Universe, the observational data intervene not in a marginal way, but with clear and unequivocal results: they confirm, by impressive agreement, the correct theories and they disprove, by equally impressive disagreement, the wrong ones. Also of this we have given a significant example.

\section*{Acknowledgments}
It is a pleasure to thank the many students who have, trough the years, collaborated with me to reach these conclusions.

%


\end{document}